\documentclass[conference]{IEEEtran}
\IEEEoverridecommandlockouts

\usepackage{cite}
\usepackage{amsmath,amssymb,amsfonts}
\usepackage{algorithm}
\usepackage{algpseudocode}
\usepackage{graphicx}
\usepackage{textcomp}
\usepackage{xcolor}

\def\BibTeX{{\rm B\kern-.05em{\sc i\kern-.025em b}\kern-.08em
    T\kern-.1667em\lower.7ex\hbox{E}\kern-.125emX}}

\newcommand{\ket}[1]{\left| #1 \right\rangle}

\begin{document}

\title{Tensor-Based Binary Graph Encoding for Variational Quantum Classifiers \\
}

\author{\IEEEauthorblockN{Shiwen An, Konstantinos Slavakis}
\IEEEauthorblockA{
    Department of Information and Communication Engineering\\
    \textit{Institute of Science Tokyo}\\
    Japan \\
an.s.aa@m.titech.ac.jp, slavakis@ict.eng.isct.ac.jp}
}


\maketitle

\begin{abstract}
    Quantum computing has been a prominent research area for decades, inspiring transformative fields such as quantum simulation, quantum teleportation, and quantum machine learning (QML), which are undergoing rapid development. Within QML, hybrid classical-quantum algorithms like Quantum Neural Networks (QNNs) and Variational Quantum Classifiers (VQCs) have shown promise in leveraging quantum circuits and classical optimizers to classify classical data efficiently.
    Simultaneously, classical machine learning has made significant strides in graph classification, employing Graph Neural Networks (GNNs) to analyze systems ranging from large-scale structures like the Large Hadron Collider to molecular and biological systems like proteins and DNA. Combining the advancements in quantum computing and graph classification presents a unique opportunity to develop quantum algorithms capable of extracting features from graphs and performing their classification effectively.
    In this paper, we propose a novel quantum encoding framework for graph classification using VQCs. Unlike existing approaches such as PCA-VQC, which rely on dimensionality reduction techniques like Principal Component Analysis (PCA) and may lead to information loss, our method preserves the integrity of graph data. Furthermore, our encoding approach is optimized for Noise-Intermediate Scale Quantum (NISQ) devices, requiring a limited number of qubits while achieving comparable or superior classification performance to PCA-VQC. By constructing slightly more complex circuits tailored for graph encoding, we demonstrate that VQCs can effectively classify graphs within the constraints of current quantum hardware.    
\end{abstract}


\section{Introduction}

Recent advancements in Quantum Computing and Quantum Machine Learning have 
positioned Variational Quantum Circuits (VQCs) as pivotal models for 
classification tasks \cite{McClean_2016}. Applications in fields such as drug discovery, 
materials engineering, and financial investment \cite{Nakaji_2022} increasingly leverage quantum algorithms due to their potential to address complex problems efficiently. Quantum computing, particularly when dealing with large datasets and intricate operations, outperforms classical approaches by utilizing quantum phenomena like superposition and entanglement. Foundational algorithms such as Grover's \cite{grover1996fastquantummechanicalalgorithm} and Shor's \cite{Shor_1997} exemplify the advantages of quantum computing.

Graph data, characterized by its rich relational structure, is 
critical in tasks such as modeling physical systems, analyzing molecular structures, 
predicting protein interactions, and classifying diseases \cite{zhou2021graphneuralnetworksreview}. 
In classical scenarios, Graph Neural Networks (GNNs) effectively process graph-structured 
data by employing message-passing mechanisms between nodes. Variants of GNNs, 
such as Graph Convolutional Networks, Graph Attention Networks, and Graph Recurrent Networks, 
have achieved groundbreaking results across diverse deep learning tasks. 
Classical methods for graph processing typically follow two mainstream approaches: 
Graph Embedding Methods and Graph Kernels. The former focuses on representing graphs as vectors of permutation-invariant features for use with standard machine learning algorithms \cite{battaglia2018relationalinductivebiasesdeep}, while the latter involves encoding graphs into high-dimensional Hilbert spaces \cite{Cosmo_2024}.

Building on these classical methodologies, quantum computing offers novel ways to process graph-structured data by leveraging quantum encodings and computational paradigms. To bridge classical and quantum domains, graph encoding methods transform classical graph data into quantum states using techniques such as Basis Encoding, Amplitude Encoding, and Angle Encoding \cite{zhao2019statepreparationbasedquantum}. These encodings provide a foundation for utilizing quantum algorithms that exploit quantum parallelism to address graph-based learning tasks more efficiently than classical methods in certain scenarios. The resulting quantum states are classified using Variational Quantum Classifiers (VQCs), circuit-centric models that iteratively optimize quantum gates based on input data and target labels \cite{Schuld_2020}. VQCs, as shown in \cite{McClean_2016}, offer a flexible and trainable architecture where parameters are optimized to minimize a cost function, enabling effective quantum state classification. These quantum state representations enable hierarchical and tensor-based architectures \cite{stoudenmire2017supervisedlearningquantuminspiredtensor}, improving scalability on quantum devices.

Previous works, such as those incorporating Graph Convolutional Network models, explore processes like aligning transitive vertices between graphs and transforming arbitrary-sized graphs into fixed-sized grid structures \cite{bai_2023}. Additional research focuses on feature extraction techniques for quantum graph processing. However, these methods often suffer from information loss during encoding or alignment.

This paper introduces a novel graph encoding regime that advances VQC design through enhanced state representation strategies, addressing the challenges of complex classification tasks in quantum environments. Our contributions are summarized as follows:
\begin{itemize}
    \item We propose a tensor-based or Ising model-inspired matrix representation for complex graph structures.
    \item The encoding method offers superior time complexity compared to traditional graph embedding approaches, enabling efficient qubit utilization. Large graphs can be processed even with limited qubits.
    \item The proposed method transforms graphs into quantum data, rather than merely quantum-encoded classical data.
\end{itemize}

The remainder of this paper is structured as follows. Section II discusses the background of quantum machine learning, with a focus on VQC methodologies and graph processing strategies like PAC-VQC. Section III formalizes the proposed quantum graph encoding method, analyzing its time and space complexities and presenting schematic views and pseudocode. Section IV provides numerical results and comparisons using small biological datasets. Finally, we conclude with a discussion of future work, with proofs of positive semi-definite matrices included in the appendix.

\section{Preliminaries}

Quantum Machine Learning leverages quantum algorithms to address computational challenges that are infeasible for classical systems. Key among QML methodologies are Variational Quantum Circuits (VQCs) and tensor network states, which provide mechanisms for encoding classical data into quantum representations and performing efficient classification tasks. This section outlines the theoretical foundations of graph encoding and quantum state representations integral to this work.

\subsection{Encoding for Quantum Data}

Data encoding is fundamental in QML, transforming classical data structures such as vectors and matrices into quantum states suitable for processing. These encodings enable quantum algorithms to handle classical inputs efficiently. Below, we discuss two prominent quantum encoding techniques.

\subsubsection{Amplitude Encoding}
Amplitude encoding is a method that maps a classical vector into 
the amplitudes of a quantum state \cite{PhysRevA.102.032420} \cite{PhysRevA.107.012422}. Given a vector \( x \) of length \( N \), it is encoded into an \( n \)-qubit quantum state where \( n = \lceil \log_2(N) \rceil \). The resulting quantum state can be expressed as:
\begin{equation}
    \ket{x} = \sum_{i=1}^N x_i \ket{i}
\end{equation}
where \( \ket{i} \) represents the computational basis states of the Hilbert space. The input vector must be normalized to ensure the quantum state's validity, satisfying \( \|x\| = 1 \). This technique efficiently encodes classical information into a quantum framework, preserving the vector's structure in the quantum domain.

\subsubsection{Angle Encoding}
Angle encoding uses the angles of rotation gates to encode classical data 
into quantum states \cite{PhysRevA.107.012422} \cite{PhysRevA.102.032420}. Each element of the input vector determines the rotation applied to a corresponding qubit. The quantum state is represented as:
\begin{equation}
    \ket{x} = \bigotimes_{i=1}^n R(x_i) \ket{0^n}
\end{equation}
where \( R(x_i) \) is a parameterized rotation gate, 
such as \( R_x \), \( R_y \), or \( R_z \). 
The dimension of the input vector dictates the number of qubits required, 
with each qubit undergoing a rotation determined by the respective 
element of the input vector. 
Angle encoding is particularly effective for translating 
feature vectors into quantum states by utilizing gate-based operations, and will 
be utilized in our Graph Encoding Regime.

\subsection{Graph Representation in Quantum Computing}

Graph encoding translates classical graph structures into quantum representations, enabling the application of quantum algorithms for graph-based data analysis. For instance, in a quantum walk framework, each vertex \( u \) of a graph \( G = (V, E) \) is associated with a quantum state, and the walker’s state vector at time \( t \) is:
\[
\ket{\psi_t} = \sum_{u \in V} \alpha_u(t) \ket{u},
\]
where \( \alpha_u(t) \) denotes the amplitude at vertex \( u \), and \( |\alpha_u(t)|^2 \) gives the probability of finding the walker at \( u \). The graph's connectivity is captured by the adjacency matrix \( A \):
\begin{equation}
    A_{uv} = \begin{cases}
        1, & \text{if } (u, v) \in E, \\
        0, & \text{otherwise}.
        \end{cases}
\end{equation}
For continuous-time quantum walks, the Hamiltonian \( H \) is typically the graph Laplacian \( L = D - A \), where \( D \) is the degree matrix. The state evolution follows Schrödinger's equation:
\[
i \frac{d}{dt} \ket{\psi} = H \ket{\psi}, \quad \ket{\psi(t)} = e^{-i H t} \ket{\psi(0)}.
\]
This formulation enables efficient quantum processing of graph data. More importantly, the graph classification task, which can be 
divided to basically two categories. One classification focuses on the node level classification \cite{liu2023surveygraphclassificationlink}, the other focus on the entire graph 
level classification. This paper will focus on the graph classification \cite{liu2023surveygraphclassificationlink}. 
 
\subsection{Variational Quantum Circuits (VQCs)} 

Variational Quantum Circuits (VQCs) are a class of quantum circuits designed with tunable parameters that are optimized to perform specific computational tasks. They have emerged as a powerful framework in quantum machine learning, offering the flexibility to adapt to various optimization problems, including classification, regression, and generative modeling \cite{Benedetti_2019}.
\begin{figure}
    \caption{Variational Quantum Circuit construction, the applied one is usually more complicated. In the instance 
    here it shows VQC in 4 qubits and 2 layers and some data encoding. \newline}
    \includegraphics[width=\linewidth]{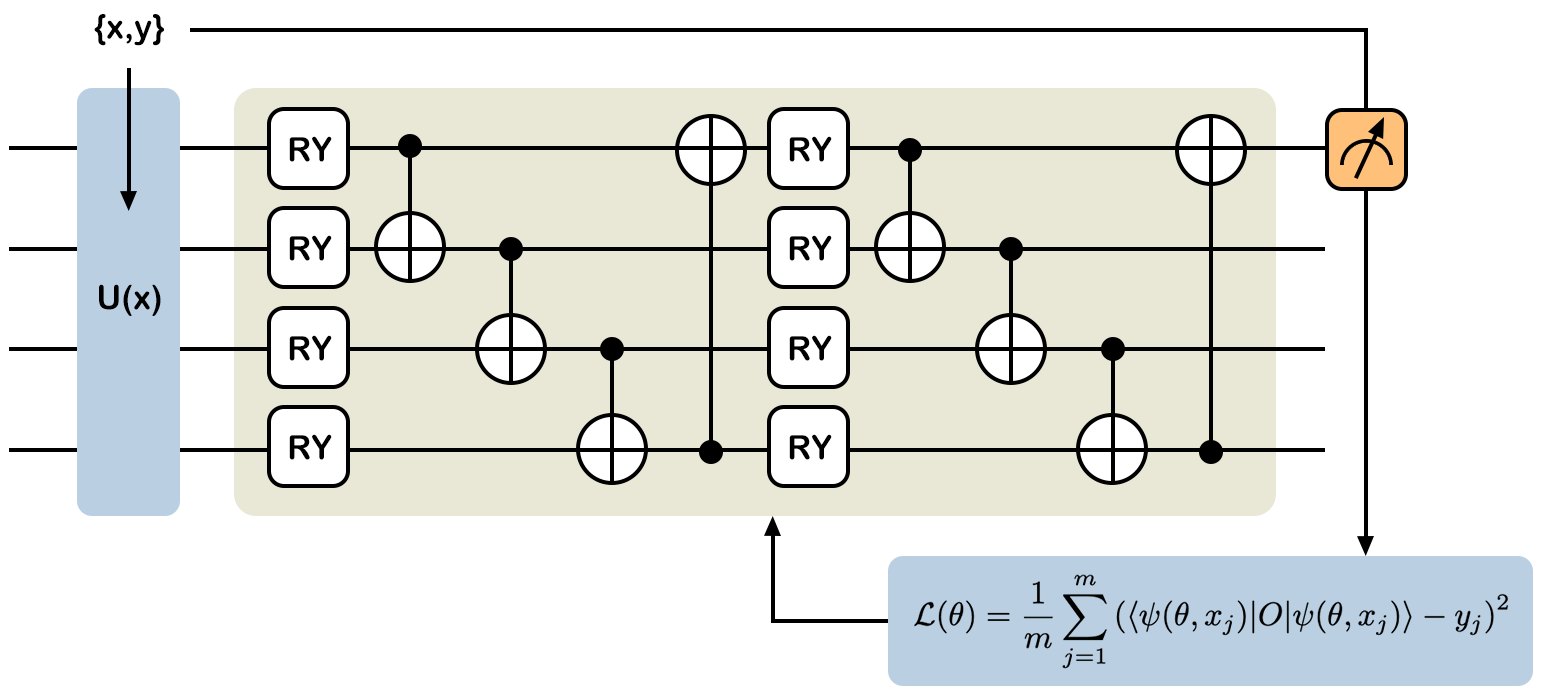}
\end{figure}
A VQC consists of three primary components: an initial state preparation layer, a parameterized ansatz layer, and a measurement layer. The initial state preparation layer encodes classical data into a quantum state, often leveraging encoding techniques such as amplitude encoding or angle encoding. The ansatz layer applies a series of parameterized quantum gates to the prepared state, introducing the tunable parameters \( \theta \). Finally, the measurement layer extracts information from the quantum state to compute the cost function for optimization.

The evolution of the quantum state in a VQC can be expressed as:
\[
\ket{\psi(\theta)} = U(\theta) \ket{\psi_{\text{enc}}},
\]
where \( \ket{\psi_{\text{enc}}} \) is the encoded quantum state, and \( U(\theta) \) represents the unitary transformation defined by the parameterized gates.
 
The optimization process in VQCs involves minimizing a cost function \( \mathcal{L}(\theta) \), which quantifies the difference between the circuit's output and the desired target. A typical cost function for classification tasks is given by:
\begin{equation}
    \mathcal{L}(\theta) = \frac{1}{m} \sum_{j=1}^m \left( \langle \psi(\theta, x_j) | O | \psi(\theta, x_j) \rangle - y_j \right)^2
\end{equation}
where \( m \) is the number of training samples, \( O \) is an observable such as the Pauli-Z operator, \( y_j \) represents the target labels, and \( \langle \psi(\theta, x_j) | O | \psi(\theta, x_j) \rangle \) corresponds to the expectation value of the observable.

Training the VQC requires an iterative process of parameter adjustment, 
typically using gradient-based optimization algorithms \cite{Schuld_2019}. 
Gradients can be computed using the parameter-shift rule \cite{Wierichs2022generalparameter}, 
which is well-suited to quantum systems. Advanced optimization techniques, 
such as the quantum natural gradient \cite{Stokes_2020}, can further accelerate convergence by considering the geometry of the parameter space.

\subsection{Principal Component Analysis \& VQC}

Principal Component Analysis (PCA) \cite{Hotelling1933} is a widely used technique in graph analysis and serves as a preprocessing step for quantum graph analysis \cite{chen2020hybridquantumclassicalclassifierbased}. Graphs are represented by their adjacency or Laplacian matrices, from which principal components are extracted via singular value decomposition (SVD). These components are encoded into quantum states using techniques such as amplitude or angle encoding, enabling their processing within the VQC framework.

\section{Graph Encoding Regime}

In this section, we introduce our new Graph Encoding Regime 
for Variational Quantum Circuits (VQC). 
The entire algorithm is referred to as Encoded Graph VQC (EG-VQC) 
in the subsequent numerical simulation section.

\subsection{Encoded Graph Representation}
In this work, we propose a graph encoding structure inspired by 
the Ising-model formalism \cite{Cervera_Lierta_2018}, 
utilizing tensor product representations. 
Specifically, we encode graph information into a 
combinatorial framework using Pauli-\(Z\) matrices and the Identity matrix, commonly 
used as quantum gate representations. 

Considering a graph \(G(V, E)\), where \(V\) represents the set of vertices, 
\( V = \{v_i| i \in \{1,\dots, N_G\} \}\), 
and \(E\) represents the set of edges, \(E = \{ (v_i,v_j) | v_i \in V, v_j \in V \}\). 
The graph can naturally be decomposed into two components: 
1. the edge representation, analogous to interactions 
in the Ising model, and 
2. the vertex representation, encoded based on vertex indices. 

Each vertex \(v_i\) in the graph, if labeled as the \(i\)-th vertex, 
is represented by its binary encoding, defined as:
\begin{equation}
    i = \sum_{k=0}^{N-1} r_k 2^k, \quad r_k \in \{0,1\},
\end{equation}
where \(N\) is the total number of qubits in the VQC or algorithm being designed. 
Using this binary representation, it follows that 
the number of vertices in the graph satisfies \(N_G < 2^N\). 
Each vertex \(v_i\) is encoded as a tensor product of Pauli-\(Z\) operators based on the 
binary encoding of the \(i\)-th vertex in the graph:
\begin{equation}
    v_i \rightarrow \bigotimes_{k=0}^{N-1} \mathcal{Z}^{r_k}_i.
    \label{eqn:vertex}
\end{equation}
For a concrete example, the first vertex in the graph \(G\), with a binary value of \(i=0001\) in a 4-qubit 
quantum computing system, is encoded as:
\[
v_1 \rightarrow \mathcal{I} \otimes \mathcal{I} \otimes \mathcal{I} \otimes \mathcal{Z}, \quad \text{where } N=4.
\]
The edges of the graph, which represent interactions between pairs of vertices \((v_i, v_j) \in E\), 
are encoded similarly by using the product of two vertices to represent their connectivity. 
The edge representation incorporates a coupling term \(J_{i,j}\) 
between the \(i\)-th and \(j\)-th vertices:
\begin{equation}
    (v_i, v_j) \rightarrow J_{i,j} \bigotimes_{k=0}^{N-1} \mathcal{Z}^{r_k}_i \bigotimes_{k=0}^{N-1} \mathcal{Z}^{r_k}_j = J_{i,j} \mathcal{Z}^\otimes_i \mathcal{Z}^\otimes_j,
    \label{eqn:edge}
\end{equation}
where \(\mathcal{Z}^\otimes_i\) is a simplified notation for the 
tensor product of Pauli-\(Z\) operators corresponding to the \(i\)-th vertex. 
The complete graph representation is expressed as the sum of contributions from all edges and vertices:
\begin{equation}
    H(G) \rightarrow \sum_{\{i,j\} \in E} J_{i,j} \mathcal{Z}^\otimes_i \mathcal{Z}^\otimes_j + \sum_{i \in V} h_i \mathcal{Z}^\otimes_i.
\end{equation}
The interaction weight \(J_{i,j}\) is derived from the normalized weight value over the entire graph. $h_i$ represents the weighted degree of the vertex. 
This normalization reparameterizes the graph to accelerate the training of deep neural networks 
\cite{salimans2016weightnormalizationsimplereparameterization}.
\begin{figure*}[ht]
    \centering
    \caption{Overview of the proposed method using the new Graph Encoding Regime and 
    VQC. The conversion of each graph into an Ising-model-like quantum circuit structure is expected to include complete interactions 
    among vertices to better formulate the prediction results.}
    \includegraphics[width=\textwidth]{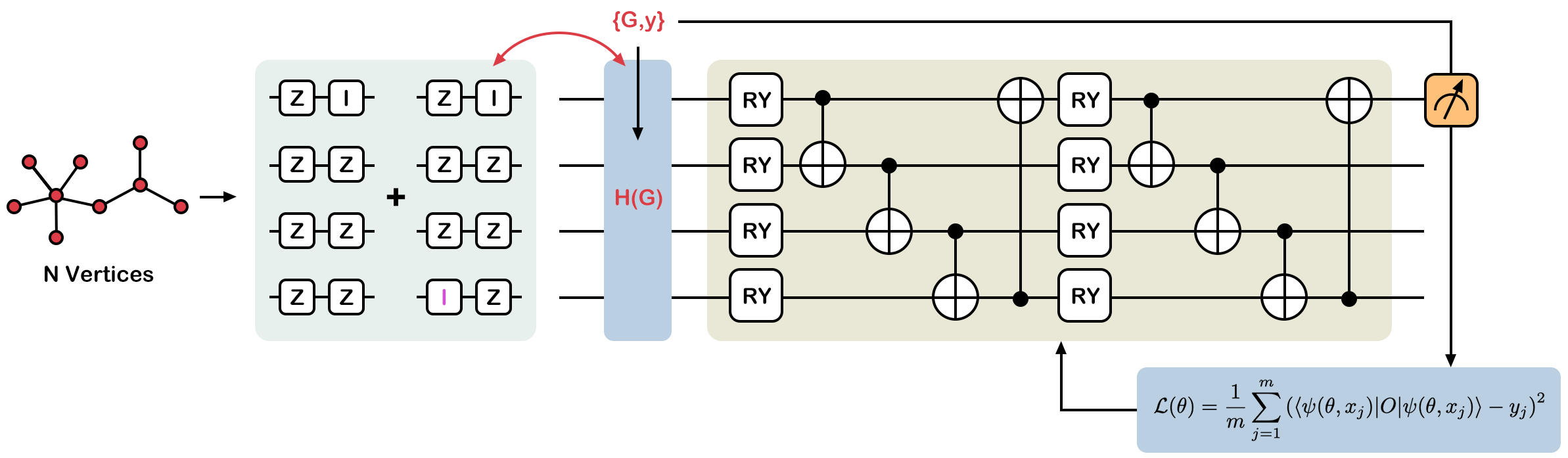}
    \label{fig:vqc_overview}
\end{figure*}

\subsubsection{Boundary Condition}
To ensure the physical consistency of the graph representation, 
the encoded Hamiltonian better satisfy the condition of eigenvalues in range of \([-1,1]\). 
This requirement guarantees compatibility with the VQC framework and cross entropy loss function, 
enabling effective use in computations and modeling tasks. 
A brief proof of this condition under special circumstances is provided in the appendix.

\subsection{Algorithmic Implementation}

The process of converting a graph \( G \) into its Graph Encoding Regime form involves 
the following algorithmic steps:
\begin{enumerate}
    \item Initialize the quantum device with \(N\) qubits, representing the system's degrees of freedom.  
    \item Encode each graph vertex as a tensor product of Pauli-\(Z\) operators, as described in Eqn.~\ref{eqn:vertex}.  
    \item Encode the edges, representing interactions between vertices, as weighted terms, as shown in Eqn.~\ref{eqn:edge}.  
\end{enumerate}
At each step, the input graph data is transformed into a quantum encoding in circuit format. The circuit is then augmented with variational layers designed to optimize performance during training. A detailed pseudocode description is provided to outline the entire procedure.

\begin{algorithm}[htbp]
    \caption{Tensor-Based Quantum Graph Encoding on VQC}
    \label{alg:vqc}
    \begin{algorithmic}[1]
    \Require Graphs, labels $(G,y)$
    \Ensure Trained quantum circuit model
    \Function{EncodeVertex}{vertex}
        \State Convert \texttt{vertex} to $N$-bit binary
        \State Map bits to Pauli operators $ \bigotimes_{k=0}^{N-1} \mathcal{Z}^{r_k}_i.$
    \EndFunction
    \Function{EncodeInteraction}{v1, v2, weight}
        \State Retrieve Pauli operators for $v1$ and $v2$
        \State Apply interaction based on $weight$ \newline $ \bigotimes_{k=0}^{N-1} \mathcal{Z}^{r_k}_i \bigotimes_{k=0}^{N-1} \mathcal{Z}^{r_k}_j = J_{i,j} \mathcal{Z}^\otimes_i \mathcal{Z}^\otimes_j$
    \EndFunction
    \State Define a quantum device with $N$ qubits
    \State Initialize qubits with Hadamard gates
    \For{each edge $(v_1, v_2, weight)$ in $G$}
        \State \Call{EncodeInteraction}{$v_1, v_2, weight$}
    \EndFor
    \State Apply variational layers with $RY(\theta)$ and $CNOT$ gate
    \State Measure qubits $ p_i = \langle 0 | H(G) U(\theta)|0\rangle$ to compute predictions
    \For{each training epoch}
        \State Compute loss using cross-entropy $\mathcal{L}$
        \State Update parameters with optimizer
    \EndFor
    \State Evaluate model performance on validation data    
    \end{algorithmic}
\end{algorithm}


\subsection{Training, Evaluation, and Complexity Analysis}

The training process optimizes a cross-entropy loss function, given by  
\(\mathcal{L} = -\frac{1}{M} \sum_{i=1}^M \left( y_i \log(p_i) + (1 - y_i) \log(1 - p_i) \right)\),  
where the probabilities \(p_i\) are obtained from VQC measurements. Optimization is performed using the Adam optimizer \cite{kingma2017adammethodstochasticoptimization} over 100 epochs \cite{Goodfellow-et-al-2016}. The dataset is divided into training and validation sets, with metrics such as validation loss and accuracy monitored at each epoch.

To evaluate the efficiency of the proposed Graph Encoding Regime, we analyze both computational and storage complexities.

\subsubsection{Computational Complexity}

The computational complexity of traditional PCA-based VQC methods is dominated by the time required for feature extraction. Performing Principal Component Analysis (PCA) using Singular Value Decomposition (SVD) typically has a time complexity of \(O(N_G^3)\) when processing the adjacency or Laplacian matrix of a graph \(G\). 

In contrast, for a graph with \(N_G\) vertices, the maximum number of possible edges is \(\frac{N_G(N_G-1)}{2}\), resulting in a time complexity of \(O(N_G^2)\) for the proposed graph encoding. This approach avoids the information loss commonly associated with PCA-based dimensionality reduction. 

Empirical results show that the proposed method significantly improves computational efficiency for larger graphs, outperforming existing methods in both speed and resource utilization.

\subsubsection{Storage Complexity}

The storage requirements for traditional PCA-based VQC methods and the proposed graph encoding regime are comparable but with important distinctions. PCA is constrained by the quantum device's capability, particularly in simulations. For instance, desktop simulators with 16GB RAM are typically limited to a 10-qubit system, corresponding to \(N_G \leq 2^N \leq 1024\) vertices. This constraint significantly slows processing and imposes strict alignment of vertices to ensure compatibility with VQC operations.

In contrast, the proposed graph encoding regime eliminates the need for strict vertex alignment. Its dynamic space utilization adapts efficiently to the structure and size of the input graph, making it more suitable for handling diverse and complex datasets. This flexibility enhances its practicality in real-world applications.

\section{Numerical Simulation}
In this section, we apply the proposed EG-VQC framework to quantum simulators and compare its performance with PCA-VQC in detail.
\subsection{Experimental Setup}
All experiments were conducted using the quantum computing library Pennylane \cite{bergholm2022pennylaneautomaticdifferentiationhybrid}, developed and supported by Xanadu. Both simulation environments and real photonic quantum backends were utilized \cite{Killoran2019strawberryfields, Bromley_2020strawberryfields}. 
To evaluate the performance of the proposed graph encoding method and its applicability in variational quantum algorithms, we used three standard graph-based datasets: MUTAG \cite{debnath2018survey}, PROTEIN \cite{Borgwardt2005}, and ENZYME \cite{Schwede2007}. The datasets are described as follows:
\begin{itemize}
    \item \textbf{MUTAG}: A dataset of mutagenic aromatic and heteroaromatic nitro compounds, represented as graphs where nodes correspond to atom types and edges represent chemical bonds \cite{debnath2018survey}.
    \item \textbf{PROTEIN}: A dataset of protein structures, with graphs representing proteins. Nodes correspond to amino acids, and edges denote their interactions. Graphs contain up to 1,000 vertices and are commonly used for classification tasks \cite{Borgwardt2005}.
    \item \textbf{ENZYME}: A dataset representing enzyme structures, where nodes represent atoms in the enzyme, and edges denote bonds between them. Graph sizes range from 30 to 100 vertices, and the dataset is often used for enzyme classification tasks \cite{Schwede2007}.
\end{itemize}

Each dataset was preprocessed to generate adjacency matrices and graph-level labels for classification tasks. For the proposed graph encoding method, classical graph data were converted into quantum gate representations suitable for VQC. Algorithms were implemented in Python using libraries such as NumPy \cite{Harris_2020} and Scikit-learn \cite{pedregosa2018scikitlearnmachinelearningpython}. Quantum circuits were simulated on the "default.qubit" backend in Pennylane \cite{bergholm2022pennylaneautomaticdifferentiationhybrid} to efficiently test the encoding scheme.

\subsection{Parameter Settings and Evaluations}

The following parameters were used for the experiments, based on different condition of the graph datasets, the necessary number of qubits, circuit depth in VQC all need to be modified:

\begin{itemize}
    \item \textbf{Number of Qubits}: \(N = \log(N_G)\) qubits were used to 
    encode graph vertices and interactions, 
    ensuring scalability for graphs of moderate size. For MUTAG and ENZYME datasets, 5 or 7 qubits are enough, but for PROTEIN dataset it may require more than 10 qubits to construct the circuit. 
    \item \textbf{Circuit Depth}: The variational quantum circuit included 3 layers of parameterized \(RY\) gates for MUTAG classification, interleaved with entangling \(CNOT\) gates. For larger graphs with more vertices, additional circuit layers (e.g., 5, 7) were introduced to accommodate increased complexity.
    \item \textbf{Training Configuration}: The Adam optimizer \cite{kingma2017adammethodstochasticoptimization} with a learning rate of \(0.01\) was used for 100 epochs, minimizing the cross-entropy loss function for graph classification.
    \item \textbf{Train-Test Split}: A 90:10 split was applied for training and testing, with stratified sampling to maintain label distribution across datasets.
\end{itemize}

\subsection{Results and Analysis}

\begin{figure}[ht]
    \includegraphics[width=\linewidth]{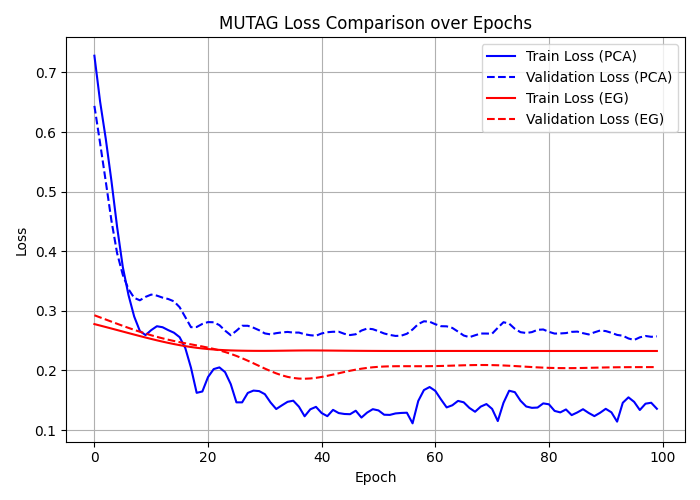}
    \includegraphics[width=\linewidth]{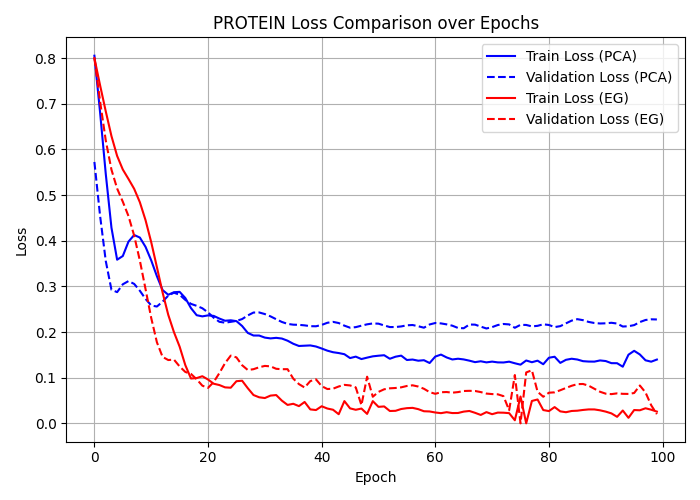}
    \caption{Comparison of loss over epochs between 
    PCA-VQC and the proposed EG-VQC method. 
    The training process using the MUTAG and PROTEIN are shown. The relative low 
    decrement of MUTAG may due to its simple graph structure compared to the others. }
\end{figure}

\subsubsection{Classification Accuracy}

The classification accuracies of EG-VQC and PCA-VQC are summarized in Table \ref{tab:accuracy_comparison}.

\begin{table}[h!]
    \centering
    \caption{Accuracy Comparison of EG-VQC and PCA-VQC}
    \begin{tabular}{|c|c|c|}
    \hline
    \textbf{Model}   & \textbf{Dataset} & \textbf{Accuracy (\%)}     \\ \hline
    PCA-VQC          & MUTAG            & \(75.1 \pm 0.2\)         \\ 
                     & PROTEIN          & \(73.6 \pm 0.2\)         \\ 
                     & ENZYME           & \(65.7 \pm 0.6\)         \\ \hline
    \textbf{EG-VQC}  & MUTAG            & \(80.3 \pm 0.3\)         \\ 
            (ours)   & PROTEIN          & \(76.8 \pm 0.4\)         \\ 
                     & ENZYME           & \(73.5 \pm 0.4\)         \\ \hline
    \end{tabular}
    \label{tab:accuracy_comparison}
\end{table}

These results highlight the effectiveness of EG-VQC in capturing structural information for graph classification tasks. While classical algorithms based on Graph Neural Networks (GNNs) may achieve higher accuracies \cite{AMOUZAD2024121280}, the proposed quantum encoding approach improves existing quantum algorithms for graph classification.

\subsubsection{Loss Trends}

The cross-entropy loss for both training and validation consistently decreased, indicating stable learning dynamics. In contrast, PCA-VQC showed signs of overfitting, as evident from higher validation losses. By encoding complete graph information, the EG-VQC approach reduces overfitting effects even without applying additional regularization techniques \cite{Goodfellow-et-al-2016}.
\section{Conclusion and Discussions}

In this paper, we demonstrated the effectiveness of Quantum Graph Encoding (EG-VQC). The integration of vertex and edge encodings facilitated efficient learning of graph-level features. The experimental results confirm the robustness of our graph encoding approach, showcasing its ability to scale effectively with graphs containing a relatively large number of vertices. 
The quantum-inspired encoding method exhibited scalability across datasets with varying sizes and complexities, offering significant advantages in graph analysis and classification tasks. The proposed approach is well-suited for integration with downstream quantum algorithms, enabling enhanced performance in graph-based computations. 
Future work will focus on extending this encoding method to handle larger and more complex datasets, as well as leveraging quantum hardware for real-time inference to further enhance its applicability in practical scenarios.

\appendix

\subsection{Eigenvalue Range \([-1, 1]\)}

We now prove that the eigenvalues of the encoded graph Hamiltonian \( H(G) \) are bounded within the range \([-1, 1]\). The Hamiltonian is given by:

\[
H(G) = \sum_{\{i,j\} \in E} J_{i,j} \mathcal{Z}^\otimes_i \mathcal{Z}^\otimes_j + \sum_{i \in V} h_i \mathcal{Z}^\otimes_i,
\]

where \( \mathcal{Z}^\otimes_i \) represents the tensor product of Pauli-\( \mathcal{Z} \) operators at node \( i \), \( J_{i,j} \) are the coupling coefficients, and \( h_i \) are node-specific coefficients.

\subsubsection{Eigenvalues of Pauli-\( \mathcal{Z} \) Operators}

Each Pauli-\( \mathcal{Z} \) operator has eigenvalues \( \pm 1 \). Consequently, for the tensor product \( \mathcal{Z}^\otimes_i \), the eigenvalues of \( \mathcal{Z}^\otimes_i \mathcal{Z}^\otimes_j \) are also \( \pm 1 \).

\subsubsection{Edge Contribution}

The edge term \( J_{i,j} \mathcal{Z}^\otimes_i \mathcal{Z}^\otimes_j \) contributes eigenvalues \( \pm J_{i,j} \), and thus the eigenvalues of the edge contribution are bounded by \( \pm |J_{i,j}| \).

\subsubsection{Node Contribution}

For the node term \( h_i \mathcal{Z}^\otimes_i \), the eigenvalues are \( \pm h_i \), implying the eigenvalues are bounded by \( \pm |h_i| \).

\subsubsection{Combined Contribution}

The total Hamiltonian \( H(G) \) is the sum of edge and node contributions. Since eigenvalues of tensor product terms \( \mathcal{Z}^\otimes_i \mathcal{Z}^\otimes_j \) are bounded by \( \pm 1 \), and the coefficients \( J_{i,j} \) and \( h_i \) are multiplied with these terms, the eigenvalues of \( H(G) \) are bounded by the maximum of the absolute values of \( J_{i,j} \) and \( h_i \).

\subsubsection{Bounding the Eigenvalues of \( H(G) \)}

Let \( \Delta_J = \max_{i,j} |J_{i,j}| \) and \( \Delta_h = \max_{i} |h_i| \). Then the eigenvalues of \( H(G) \) are bounded by:

\[
\lambda(H(G)) \in \left[ -(\Delta_J + \Delta_h), \Delta_J + \Delta_h \right].
\]

\subsubsection{Conclusion}

If \( \Delta_J + \Delta_h\leq 1 \), the eigenvalues of \( H(G) \) are within the range:
\(
\lambda(H(G)) \in [-1, 1].
\)

\bibliographystyle{IEEEtran}
\bibliography{ref}

\begin{thebibliography}{10}
\providecommand{\url}[1]{#1}
\csname url@samestyle\endcsname
\providecommand{\newblock}{\relax}
\providecommand{\bibinfo}[2]{#2}
\providecommand{\BIBentrySTDinterwordspacing}{\spaceskip=0pt\relax}
\providecommand{\BIBentryALTinterwordstretchfactor}{4}
\providecommand{\BIBentryALTinterwordspacing}{\spaceskip=\fontdimen2\font plus
\BIBentryALTinterwordstretchfactor\fontdimen3\font minus \fontdimen4\font\relax}
\providecommand{\BIBforeignlanguage}[2]{{%
\expandafter\ifx\csname l@#1\endcsname\relax
\typeout{** WARNING: IEEEtran.bst: No hyphenation pattern has been}%
\typeout{** loaded for the language `#1'. Using the pattern for}%
\typeout{** the default language instead.}%
\else
\language=\csname l@#1\endcsname
\fi
#2}}
\providecommand{\BIBdecl}{\relax}
\BIBdecl

\bibitem{McClean_2016}
\BIBentryALTinterwordspacing
J.~R. McClean, J.~Romero, R.~Babbush, and A.~Aspuru-Guzik, ``The theory of variational hybrid quantum-classical algorithms,'' \emph{New Journal of Physics}, vol.~18, no.~2, p. 023023, Feb. 2016. [Online]. Available: \url{http://dx.doi.org/10.1088/1367-2630/18/2/023023}
\BIBentrySTDinterwordspacing

\bibitem{Nakaji_2022}
\BIBentryALTinterwordspacing
K.~Nakaji, S.~Uno, Y.~Suzuki, R.~Raymond, T.~Onodera, T.~Tanaka, H.~Tezuka, N.~Mitsuda, and N.~Yamamoto, ``Approximate amplitude encoding in shallow parameterized quantum circuits and its application to financial market indicators,'' \emph{Physical Review Research}, vol.~4, no.~2, May 2022. [Online]. Available: \url{http://dx.doi.org/10.1103/PhysRevResearch.4.023136}
\BIBentrySTDinterwordspacing

\bibitem{grover1996fastquantummechanicalalgorithm}
\BIBentryALTinterwordspacing
L.~K. Grover, ``A fast quantum mechanical algorithm for database search,'' 1996. [Online]. Available: \url{https://arxiv.org/abs/quant-ph/9605043}
\BIBentrySTDinterwordspacing

\bibitem{Shor_1997}
\BIBentryALTinterwordspacing
P.~W. Shor, ``Polynomial-time algorithms for prime factorization and discrete logarithms on a quantum computer,'' \emph{SIAM Journal on Computing}, vol.~26, no.~5, pp. 1484--1509, Oct. 1997. [Online]. Available: \url{http://dx.doi.org/10.1137/S0097539795293172}
\BIBentrySTDinterwordspacing

\bibitem{zhou2021graphneuralnetworksreview}
\BIBentryALTinterwordspacing
J.~Zhou, G.~Cui, S.~Hu, Z.~Zhang, C.~Yang, Z.~Liu, L.~Wang, C.~Li, and M.~Sun, ``Graph neural networks: A review of methods and applications,'' 2021. [Online]. Available: \url{https://arxiv.org/abs/1812.08434}
\BIBentrySTDinterwordspacing

\bibitem{battaglia2018relationalinductivebiasesdeep}
\BIBentryALTinterwordspacing
P.~W. Battaglia, J.~B. Hamrick, V.~Bapst, A.~Sanchez-Gonzalez, V.~Zambaldi, M.~Malinowski, A.~Tacchetti, D.~Raposo, A.~Santoro, R.~Faulkner, C.~Gulcehre, F.~Song, A.~Ballard, J.~Gilmer, G.~Dahl, A.~Vaswani, K.~Allen, C.~Nash, V.~Langston, C.~Dyer, N.~Heess, D.~Wierstra, P.~Kohli, M.~Botvinick, O.~Vinyals, Y.~Li, and R.~Pascanu, ``Relational inductive biases, deep learning, and graph networks,'' 2018. [Online]. Available: \url{https://arxiv.org/abs/1806.01261}
\BIBentrySTDinterwordspacing

\bibitem{Cosmo_2024}
\BIBentryALTinterwordspacing
L.~Cosmo, G.~Minello, A.~Bicciato, M.~M. Bronstein, E.~Rodolà, L.~Rossi, and A.~Torsello, ``Graph kernel neural networks,'' \emph{IEEE Transactions on Neural Networks and Learning Systems}, pp. 1--14, 2024. [Online]. Available: \url{http://dx.doi.org/10.1109/TNNLS.2024.3400850}
\BIBentrySTDinterwordspacing

\bibitem{zhao2019statepreparationbasedquantum}
\BIBentryALTinterwordspacing
J.~Zhao, Y.-C. Wu, G.-C. Guo, and G.-P. Guo, ``State preparation based on quantum phase estimation,'' 2019. [Online]. Available: \url{https://arxiv.org/abs/1912.05335}
\BIBentrySTDinterwordspacing

\bibitem{Schuld_2020}
\BIBentryALTinterwordspacing
M.~Schuld, A.~Bocharov, K.~M. Svore, and N.~Wiebe, ``Circuit-centric quantum classifiers,'' \emph{Physical Review A}, vol. 101, no.~3, Mar. 2020. [Online]. Available: \url{http://dx.doi.org/10.1103/PhysRevA.101.032308}
\BIBentrySTDinterwordspacing

\bibitem{stoudenmire2017supervisedlearningquantuminspiredtensor}
\BIBentryALTinterwordspacing
E.~M. Stoudenmire and D.~J. Schwab, ``Supervised learning with quantum-inspired tensor networks,'' 2017. [Online]. Available: \url{https://arxiv.org/abs/1605.05775}
\BIBentrySTDinterwordspacing

\bibitem{bai_2023}
L.~Bai, Y.~Jiao, L.~Cui, L.~Rossi, Y.~Wang, P.~S. Yu, and E.~R. Hancock, ``Learning graph convolutional networks based on quantum vertex information propagation,'' \emph{IEEE Transactions on Knowledge and Data Engineering}, vol.~35, no.~2, pp. 1747--1760, 2023.

\bibitem{PhysRevA.102.032420}
\BIBentryALTinterwordspacing
R.~LaRose and B.~Coyle, ``Robust data encodings for quantum classifiers,'' \emph{Phys. Rev. A}, vol. 102, p. 032420, Sep 2020. [Online]. Available: \url{https://link.aps.org/doi/10.1103/PhysRevA.102.032420}
\BIBentrySTDinterwordspacing

\bibitem{PhysRevA.107.012422}
\BIBentryALTinterwordspacing
S.~Shin, Y.~S. Teo, and H.~Jeong, ``Exponential data encoding for quantum supervised learning,'' \emph{Phys. Rev. A}, vol. 107, p. 012422, Jan 2023. [Online]. Available: \url{https://link.aps.org/doi/10.1103/PhysRevA.107.012422}
\BIBentrySTDinterwordspacing

\bibitem{liu2023surveygraphclassificationlink}
\BIBentryALTinterwordspacing
X.~Liu, J.~Chen, and Q.~Wen, ``A survey on graph classification and link prediction based on gnn,'' 2023. [Online]. Available: \url{https://arxiv.org/abs/2307.00865}
\BIBentrySTDinterwordspacing

\bibitem{Benedetti_2019}
\BIBentryALTinterwordspacing
M.~Benedetti, E.~Lloyd, S.~Sack, and M.~Fiorentini, ``Parameterized quantum circuits as machine learning models,'' \emph{Quantum Science and Technology}, vol.~4, no.~4, p. 043001, Nov. 2019. [Online]. Available: \url{http://dx.doi.org/10.1088/2058-9565/ab4eb5}
\BIBentrySTDinterwordspacing

\bibitem{Schuld_2019}
\BIBentryALTinterwordspacing
M.~Schuld, V.~Bergholm, C.~Gogolin, J.~Izaac, and N.~Killoran, ``Evaluating analytic gradients on quantum hardware,'' \emph{Physical Review A}, vol.~99, no.~3, Mar. 2019. [Online]. Available: \url{http://dx.doi.org/10.1103/PhysRevA.99.032331}
\BIBentrySTDinterwordspacing

\bibitem{Wierichs2022generalparameter}
\BIBentryALTinterwordspacing
D.~Wierichs, J.~Izaac, C.~Wang, and C.~Y.-Y. Lin, ``General parameter-shift rules for quantum gradients,'' \emph{{Quantum}}, vol.~6, p. 677, Mar. 2022. [Online]. Available: \url{https://doi.org/10.22331/q-2022-03-30-677}
\BIBentrySTDinterwordspacing

\bibitem{Stokes_2020}
\BIBentryALTinterwordspacing
J.~Stokes, J.~Izaac, N.~Killoran, and G.~Carleo, ``Quantum natural gradient,'' \emph{Quantum}, vol.~4, p. 269, May 2020. [Online]. Available: \url{http://dx.doi.org/10.22331/q-2020-05-25-269}
\BIBentrySTDinterwordspacing

\bibitem{Hotelling1933}
\BIBentryALTinterwordspacing
H.~Hotelling, ``Analysis of a complex of statistical variables into principal components,'' \emph{Journal of Educational Psychology}, vol.~24, no.~6, pp. 417--441, 1933. [Online]. Available: \url{https://doi.org/10.1037/h0071325}
\BIBentrySTDinterwordspacing

\bibitem{chen2020hybridquantumclassicalclassifierbased}
\BIBentryALTinterwordspacing
S.~Y.-C. Chen, C.-M. Huang, C.-W. Hsing, and Y.-J. Kao, ``Hybrid quantum-classical classifier based on tensor network and variational quantum circuit,'' 2020. [Online]. Available: \url{https://arxiv.org/abs/2011.14651}
\BIBentrySTDinterwordspacing

\bibitem{Cervera_Lierta_2018}
\BIBentryALTinterwordspacing
A.~Cervera-Lierta, ``Exact ising model simulation on a quantum computer,'' \emph{Quantum}, vol.~2, p. 114, Dec. 2018. [Online]. Available: \url{http://dx.doi.org/10.22331/q-2018-12-21-114}
\BIBentrySTDinterwordspacing

\bibitem{salimans2016weightnormalizationsimplereparameterization}
\BIBentryALTinterwordspacing
T.~Salimans and D.~P. Kingma, ``Weight normalization: A simple reparameterization to accelerate training of deep neural networks,'' 2016. [Online]. Available: \url{https://arxiv.org/abs/1602.07868}
\BIBentrySTDinterwordspacing

\bibitem{kingma2017adammethodstochasticoptimization}
\BIBentryALTinterwordspacing
D.~P. Kingma and J.~Ba, ``Adam: A method for stochastic optimization,'' 2017. [Online]. Available: \url{https://arxiv.org/abs/1412.6980}
\BIBentrySTDinterwordspacing

\bibitem{Goodfellow-et-al-2016}
I.~Goodfellow, Y.~Bengio, and A.~Courville, \emph{Deep Learning}.\hskip 1em plus 0.5em minus 0.4em\relax MIT Press, 2016, \url{http://www.deeplearningbook.org}.

\bibitem{bergholm2022pennylaneautomaticdifferentiationhybrid}
\BIBentryALTinterwordspacing
V.~B. et. al, ``Pennylane: Automatic differentiation of hybrid quantum-classical computations,'' 2022. [Online]. Available: \url{https://arxiv.org/abs/1811.04968}
\BIBentrySTDinterwordspacing

\bibitem{Killoran2019strawberryfields}
\BIBentryALTinterwordspacing
N.~Killoran, J.~Izaac, N.~Quesada, V.~Bergholm, M.~Amy, and C.~Weedbrook, ``Strawberry {F}ields: {A} {S}oftware {P}latform for {P}hotonic {Q}uantum {C}omputing,'' \emph{{Quantum}}, vol.~3, p. 129, Mar. 2019. [Online]. Available: \url{https://doi.org/10.22331/q-2019-03-11-129}
\BIBentrySTDinterwordspacing

\bibitem{Bromley_2020strawberryfields}
\BIBentryALTinterwordspacing
T.~R. Bromley, J.~M. Arrazola, S.~Jahangiri, J.~Izaac, N.~Quesada, A.~D. Gran, M.~Schuld, J.~Swinarton, Z.~Zabaneh, and N.~Killoran, ``Applications of near-term photonic quantum computers: software and algorithms,'' \emph{Quantum Science and Technology}, vol.~5, no.~3, p. 034010, may 2020. [Online]. Available: \url{https://dx.doi.org/10.1088/2058-9565/ab8504}
\BIBentrySTDinterwordspacing

\bibitem{debnath2018survey}
A.~Debnath, S.~Soni, and C.~K. Reddy, ``A survey on graph neural networks,'' \emph{Journal of Machine Learning Research}, vol.~19, no.~1, pp. 99--124, 2018.

\bibitem{Borgwardt2005}
K.~M. Borgwardt, N.~Krämer, R.~Si, D.~Hans, and B.~Schölkopf, ``Protein classification with graph kernels,'' \emph{Bioinformatics}, vol.~21, no.~1, pp. 105--116, 2005.

\bibitem{Schwede2007}
T.~Schwede, J.~Kopp, G.~Benoit, F.~Kepes, W.~Thiel, G.~Deleage, and F.~L{\"o}hr, ``Enzyme classification in terms of graph-based representations,'' \emph{Journal of Computational Chemistry}, vol.~28, no.~2, pp. 292--301, 2007.

\bibitem{Harris_2020}
\BIBentryALTinterwordspacing
C.~R. Harris, K.~J. Millman, S.~J. van~der Walt, R.~Gommers, P.~Virtanen, D.~Cournapeau, E.~Wieser, J.~Taylor, S.~Berg, N.~J. Smith, R.~Kern, M.~Picus, S.~Hoyer, M.~H. van Kerkwijk, M.~Brett, A.~Haldane, J.~F. del Río, M.~Wiebe, P.~Peterson, P.~Gérard-Marchant, K.~Sheppard, T.~Reddy, W.~Weckesser, H.~Abbasi, C.~Gohlke, and T.~E. Oliphant, ``Array programming with numpy,'' \emph{Nature}, vol. 585, no. 7825, p. 357–362, Sep. 2020. [Online]. Available: \url{http://dx.doi.org/10.1038/s41586-020-2649-2}
\BIBentrySTDinterwordspacing

\bibitem{pedregosa2018scikitlearnmachinelearningpython}
\BIBentryALTinterwordspacing
F.~Pedregosa, G.~Varoquaux, A.~Gramfort, V.~Michel, B.~Thirion, O.~Grisel, M.~Blondel, A.~Müller, J.~Nothman, G.~Louppe, P.~Prettenhofer, R.~Weiss, V.~Dubourg, J.~Vanderplas, A.~Passos, D.~Cournapeau, M.~Brucher, M.~Perrot, and Édouard Duchesnay, ``Scikit-learn: Machine learning in python,'' 2018. [Online]. Available: \url{https://arxiv.org/abs/1201.0490}
\BIBentrySTDinterwordspacing

\bibitem{AMOUZAD2024121280}
\BIBentryALTinterwordspacing
A.~Amouzad, Z.~Dehghanian, S.~Saravani, M.~Amirmazlaghani, and B.~Roshanfekr, ``Graph isomorphism u-net,'' \emph{Expert Systems with Applications}, vol. 236, p. 121280, 2024. [Online]. Available: \url{https://www.sciencedirect.com/science/article/pii/S0957417423017827}
\BIBentrySTDinterwordspacing

\end{thebibliography}

\vspace{12pt}
\color{red}

\end{document}